# Deuteron and Anti-deuteron Production in $e^+e^-$ Collisions at the Z Resonance

ALEPH Collaboration

**Abstract**

Deuteron and anti-deuteron production in Z decays has been observed in the ALEPH experiment at LEP. The production rate of anti-deuterons is measured to be $(5.9 \pm 1.8 \pm 0.5) \times 10^{-6}$ per hadronic Z decay in the anti-deuteron momentum range from 0.62 to 1.03 GeV/c. The coalescence parameter $B_2$, which characterizes the likelihood of anti-deuteron production, is measured to be $0.0033 \pm 0.0013$ GeV$^2$ in Z decays. These measurements indicate that the production of anti-deuterons is suppressed in $e^+e^-$ collisions compared to that in pp and photoproduction collisions.

# The ALEPH Collaboration

S. Schael,

  *Physikalisches Institut das RWTH-Aachen, D-52056 Aachen, Germany*

R. Barate, R. Brunelière, I. De Bonis, D. Decamp, C. Goy, S. Jézéquel, J.-P. Lees, F. Martin, E. Merle,
M.-N. Minard, B. Pietrzyk, B. Trocmé

  *Laboratoire de Physique des Particules (LAPP), IN$^2$P$^3$-CNRS, F-74019 Annecy-le-Vieux Cedex, France*

S. Bravo, M.P. Casado, M. Chmeissani, J.M. Crespo, E. Fernandez, M. Fernandez-Bosman, Ll. Garrido,[15]
M. Martinez, A. Pacheco, H. Ruiz

  *Institut de Física d'Altes Energies, Universitat Autònoma de Barcelona, E-08193 Bellaterra (Barcelona), Spain*[7]

A. Colaleo, D. Creanza, N. De Filippis, M. de Palma, G. Iaselli, G. Maggi, M. Maggi, S. Nuzzo, A. Ranieri,
G. Raso,[24] F. Ruggieri, G. Selvaggi, L. Silvestris, P. Tempesta, A. Tricomi,[3] G. Zito

  *Dipartimento di Fisica, INFN Sezione di Bari, I-70126 Bari, Italy*

X. Huang, J. Lin, Q. Ouyang, T. Wang, Y. Xie, R. Xu, S. Xue, J. Zhang, L. Zhang, W. Zhao

  *Institute of High Energy Physics, Academia Sinica, Beijing, The People's Republic of China*[8]

D. Abbaneo, T. Barklow,[26] O. Buchmüller,[26] M. Cattaneo, B. Clerbaux,[23] H. Drevermann, R.W. Forty, M. Frank,
F. Gianotti, J.B. Hansen, J. Harvey, D.E. Hutchcroft,[30], P. Janot, B. Jost, M. Kado,[2] P. Mato, A. Moutoussi,
F. Ranjard, L. Rolandi, D. Schlatter, F. Teubert, A. Valassi, I. Videau

  *European Laboratory for Particle Physics (CERN), CH-1211 Geneva 23, Switzerland*

F. Badaud, S. Dessagne, A. Falvard,[20] D. Fayolle, P. Gay, J. Jousset, B. Michel, S. Monteil, D. Pallin,
J.M. Pascolo, P. Perret

  *Laboratoire de Physique Corpusculaire, Université Blaise Pascal, IN$^2$P$^3$-CNRS, Clermont-Ferrand, F-63177 Aubière, France*

J.D. Hansen, J.R. Hansen, P.H. Hansen, A.C. Kraan, B.S. Nilsson

  *Niels Bohr Institute, 2100 Copenhagen, DK-Denmark*[9]

A. Kyriakis, C. Markou, E. Simopoulou, A. Vayaki, K. Zachariadou

  *Nuclear Research Center Demokritos (NRCD), GR-15310 Attiki, Greece*

A. Blondel,[12] J.-C. Brient, F. Machefert, A. Rougé, H. Videau

  *Laoratoire Leprince-Ringuet, Ecole Polytechnique, IN$^2$P$^3$-CNRS, F-91128 Palaiseau Cedex, France*

V. Ciulli, E. Focardi, G. Parrini

  *Dipartimento di Fisica, Università di Firenze, INFN Sezione di Firenze, I-50125 Firenze, Italy*

A. Antonelli, M. Antonelli, G. Bencivenni, F. Bossi, G. Capon, F. Cerutti, V. Chiarella, P. Laurelli,
G. Mannocchi,[5] G.P. Murtas, L. Passalacqua

  *Laboratori Nazionali dell'INFN (LNF-INFN), I-00044 Frascati, Italy*

J. Kennedy, J.G. Lynch, P. Negus, V. O'Shea, A.S. Thompson

  *Department of Physics and Astronomy, University of Glasgow, Glasgow G12 8QQ,United Kingdom*[10]

S. Wasserbaech

  *Utah Valley State College, Orem, UT 84058, U.S.A.*

R. Cavanaugh,[4] S. Dhamotharan,[21] C. Geweniger, P. Hanke, V. Hepp, E.E. Kluge, A. Putzer, H. Stenzel, K. Tittel,
M. Wunsch[19]

  *Kirchhoff-Institut für Physik, Universität Heidelberg, D-69120 Heidelberg, Germany*[16]

R. Beuselinck, W. Cameron, G. Davies, P.J. Dornan, M. Girone,[1] N. Marinelli, J. Nowell, S.A. Rutherford, J.K. Sedgbeer, J.C. Thompson,[14] R. White

*Department of Physics, Imperial College, London SW7 2BZ, United Kingdom*[10]

V.M. Ghete, P. Girtler, E. Kneringer, D. Kuhn, G. Rudolph

*Institut für Experimentalphysik, Universität Innsbruck, A-6020 Innsbruck, Austria*[18]

E. Bouhova-Thacker, C.K. Bowdery, D.P. Clarke, G. Ellis, A.J. Finch, F. Foster, G. Hughes, R.W.L. Jones, M.R. Pearson, N.A. Robertson, T. Sloan, M. Smizanska

*Department of Physics, University of Lancaster, Lancaster LA1 4YB, United Kingdom*[10]

O. van der Aa, C. Delaere,[28] G.Leibenguth,[31] V. Lemaitre[29]

*Institut de Physique Nucléaire, Département de Physique, Université Catholique de Louvain, 1348 Louvain-la-Neuve, Belgium*

U. Blumenschein, F. Hölldorfer, K. Jakobs, F. Kayser, A.-S. Müller, B. Renk, H.-G. Sander, S. Schmeling, H. Wachsmuth, C. Zeitnitz, T. Ziegler

*Institut für Physik, Universität Mainz, D-55099 Mainz, Germany*[16]

A. Bonissent, P. Coyle, C. Curtil, A. Ealet, D. Fouchez, P. Payre, A. Tilquin

*Centre de Physique des Particules de Marseille, Univ Méditerranée, IN$^3$-CNRS, F-13288 Marseille, France*

F. Ragusa

*Dipartimento di Fisica, Università di Milano e INFN Sezione di Milano, I-20133 Milano, Italy.*

A. David, H. Dietl,[32] G. Ganis,[27] K. Hüttmann, G. Lütjens, W. Männer[32], H.-G. Moser, R. Settles, M. Villegas, G. Wolf

*Max-Planck-Institut für Physik, Werner-Heisenberg-Institut, D-80805 München, Germany*[16]

J. Boucrot, O. Callot, M. Davier, L. Duflot, J.-F. Grivaz, Ph. Heusse, A. Jacholkowska,[6] L. Serin, J.-J. Veillet

*Laboratoire de l'Accélérateur Linéaire, Université de Paris-Sud, IN$^8$-CNRS, F-91898 Orsay Cedex, France*

P. Azzurri, G. Bagliesi, T. Boccali, L. Foà, A. Giammanco, A. Giassi, F. Ligabue, A. Messineo, F. Palla, G. Sanguinetti, A. Sciabà, G. Sguazzoni, P. Spagnolo, R. Tenchini, A. Venturi, P.G. Verdini

*Dipartimento di Fisica dell'Università, INFN Sezione di Pisa, e Scuola Normale Superiore, I-56010 Pisa, Italy*

O. Awunor, G.A. Blair, G. Cowan, A. Garcia-Bellido, M.G. Green, T. Medcalf,[25] A. Misiejuk, J.A. Strong, P. Teixeira-Dias

*Department of Physics, Royal Holloway & Bedford New College, University of London, Egham, Surrey TW20 OEX, United Kingdom*[10]

R.W. Clifft, T.R. Edgecock, P.R. Norton, I.R. Tomalin, J.J. Ward

*Particle Physics Dept., Rutherford Appleton Laboratory, Chilton, Didcot, Oxon OX11 OQX, United Kingdom*[10]

B. Bloch-Devaux, D. Boumediene, P. Colas, B. Fabbro, E. Lançon, M.-C. Lemaire, E. Locci, P. Perez, J. Rander, B. Tuchming, B. Vallage

*CEA, DAPNIA/Service de Physique des Particules, CE-Saclay, F-91191 Gif-sur-Yvette Cedex, France*[17]

A.M. Litke, G. Taylor

*Institute for Particle Physics, University of California at Santa Cruz, Santa Cruz, CA 95064, USA*[22]

C.N. Booth, S. Cartwright, F. Combley,[25] P.N. Hodgson, M. Lehto, L.F. Thompson

*Department of Physics, University of Sheffield, Sheffield S3 7RH, United Kingdom*[10]

A. Böhrer, S. Brandt, C. Grupen, J. Hess, A. Ngac, G. Prange

*Fachbereich Physik, Universität Siegen, D-57068 Siegen, Germany*[16]

C. Borean, G. Giannini

*Dipartimento di Fisica, Università di Trieste e INFN Sezione di Trieste, I-34127 Trieste, Italy*

H. He, J. Putz, J. Rothberg

*Experimental Elementary Particle Physics, University of Washington, Seattle, WA 98195 U.S.A.*

S.R. Armstrong, K. Berkelman, K. Cranmer, D.P.S. Ferguson, Y. Gao,[13] S. González, O.J. Hayes, H. Hu, S. Jin, J. Kile, P.A. McNamara III, J. Nielsen, Y.B. Pan, J.H. von Wimmersperg-Toeller, W. Wiedenmann, J. Wu, Sau Lan Wu, X. Wu, G. Zobernig

*Department of Physics, University of Wisconsin, Madison, WI 53706, USA*[11]

G. Dissertori

*Institute for Particle Physics, ETH Hönggerberg, 8093 Zürich, Switzerland.*

---

[1] Also at CERN, 1211 Geneva 23, Switzerland.
[2] Now at Fermilab, PO Box 500, MS 352, Batavia, IL 60510, USA
[3] Also at Dipartimento di Fisica di Catania and INFN Sezione di Catania, 95129 Catania, Italy.
[4] Now at University of Florida, Department of Physics, Gainesville, Florida 32611-8440, USA
[5] Also IFSI sezione di Torino, INAF, Italy.
[6] Also at Groupe d'Astroparticules de Montpellier, Université de Montpellier II, 34095, Montpellier, France.
[7] Supported by CICYT, Spain.
[8] Supported by the National Science Foundation of China.
[9] Supported by the Danish Natural Science Research Council.
[10] Supported by the UK Particle Physics and Astronomy Research Council.
[11] Supported by the US Department of Energy, grant DE-FG0295-ER40896.
[12] Now at Departement de Physique Corpusculaire, Université de Genève, 1211 Genève 4, Switzerland.
[13] Also at Department of Physics, Tsinghua University, Beijing, The People's Republic of China.
[14] Supported by the Leverhulme Trust.
[15] Permanent address: Universitat de Barcelona, 08208 Barcelona, Spain.
[16] Supported by Bundesministerium für Bildung und Forschung, Germany.
[17] Supported by the Direction des Sciences de la Matière, C.E.A.
[18] Supported by the Austrian Ministry for Science and Transport.
[19] Now at SAP AG, 69185 Walldorf, Germany
[20] Now at Groupe d' Astroparticules de Montpellier, Université de Montpellier II, 34095 Montpellier, France.
[21] Now at BNP Paribas, 60325 Frankfurt am Mainz, Germany
[22] Supported by the US Department of Energy, grant DE-FG03-92ER40689.
[23] Now at Institut Inter-universitaire des hautes Energies (IIHE), CP 230, Université Libre de Bruxelles, 1050 Bruxelles, Belgique
[24] Now at Dipartimento di Fisica e Tecnologie Relative, Università di Palermo, Palermo, Italy.
[25] Deceased.
[26] Now at SLAC, Stanford, CA 94309, U.S.A
[27] Now at CERN, 1211 Geneva 23, Switzerland
[28] Research Fellow of the Belgium FNRS
[29] Research Associate of the Belgium FNRS
[30] Now at Liverpool University, Liverpool L69 7ZE, United Kingdom
[31] Supported by the Federal Office for Scientific, Technical and Cultural Affairs through the Interuniversity Attraction Pole P5/27
[32] Now at Henryk Niewodniczanski Institute of Nuclear Physics, Polish Academy of Sciences, Cracow, Poland

# 1 Introduction

The production of nuclei in particle collisions can be described in terms of the coalescence model [1] in which baryons produced in the quark fragmentation process coalesce into nuclei. In this model, assuming that the baryons are uncorrelated, the cross section, $\sigma_A$, for the formation of a nucleus with $A$ nucleons with total energy $E_A$ and momentum $P$, is related to that for the production of free nucleons, $\sigma_N$, with energy $E_N$ and momentum $p = P/A$, by

$$\frac{1}{\sigma}\frac{E_A \mathrm{d}^3\sigma_A}{\mathrm{d}^3 P} = B_A \left(\frac{1}{\sigma}\frac{E_N \mathrm{d}^3\sigma_N}{\mathrm{d}^3 p}\right)^A, \tag{1}$$

where $B_A$ is the coalescence parameter and $\sigma$ is the total interaction cross section. For deuteron and anti-deuteron production $A = 2$.

Deuteron and anti-deuteron production has been measured previously in heavy ion collisions [2–11], proton-proton collisions [12–15], proton-nucleus collisions [16–18] and in photoproduction [19]. The values of $B_2$ are found to be similar in photoproduction, proton-proton and proton-nucleus interactions at a value of $B_2 \sim 0.02$ GeV$^2$ [19]. In contrast, in $e^+e^-$ annihilation deuteron and anti-deuteron production seems to be suppressed [20, 21]. In the ARGUS experiment [20], in the continuum away from the $\Upsilon$ resonances, a limit on the rate of anti-deuteron production at 90% confidence level was set at $1.7 \times 10^{-5}$ per annihilation event. In Z decays the limit at 95% confidence level was found to be $0.8 \times 10^{-5}$ per hadronic Z decay by the OPAL Collaboration [21] in the momentum range from 0.35 to 1.1 GeV/c from a fraction of the available LEP data. This corresponds to a value of $B_2 < 0.003$ GeV$^2$.

In $e^+e^-$ collisions, it has been suggested that, in the string fragmentation model, such suppression is caused by correlations between the baryons in an $e^+e^-$ collision event [22]. The purpose of this paper is to measure the rate of deuteron and anti-deuteron production using the full luminosity available for the study of Z decays at LEP, in order to test the predictions from this model. This is of topical interest since it has been postulated that the production of the possibly observed pentaquark states is governed by a similar coalescence process as that for deuteron and anti-deuteron production [23].

In this paper the observation of deuterons and anti-deuterons is described from a sample of $4.07 \times 10^6$ hadronic Z decays collected by the ALEPH experiment in the years 1990-95. In Section 2 the apparatus, trigger, event and track selection procedures are described. The method of isolating the deuteron and anti-deuterons using the measured specific ionization energy loss, $\mathrm{d}E/\mathrm{d}x$, and track momenta is also described. In Section 3 the measurements made using the anti-deuteron sample are described and these are discussed in Section 4.

# 2 The Apparatus and the Selection of Events and Tracks

## 2.1 The ALEPH Detector

Collisions of positrons with electrons at LEP around the Z resonance energy were detected in the ALEPH detector which is described in detail elsewhere [24]. The components of the detector most relevant to this analysis were the tracking and trigger systems. The tracking system consisted of a silicon vertex detector, a drift chamber and a large time projection chamber in a 1.5 T axial magnetic field produced by a super-conducting coil. The silicon vertex detector (VDET) [25] provided precise track measurements very close to the interaction point. The spatial resolution for the $r\phi$ and $z$ projections (transverse to and along the beam axis, respectively) was $12\mu$m at normal incidence. The vertex detector was surrounded by a multilayer axial-wire cylindrical drift chamber, the inner tracking chamber (ITC), which was 200 cm long and measured the $r\phi$ positions of tracks at 8 radii between 16 and 26 cm. The average resolution in the $r\phi$ coordinate was 150 $\mu$m. The time projection chamber (TPC) was the main tracking detector. It was 440 cm long and provided up to 21 three dimensional space coordinates and 338 samples of ionization loss ($\mathrm{d}E/\mathrm{d}x$) for tracks at radii between 30 and 180 cm. Azimuthal ($r\phi$) and longitudinal ($z$) coordinate resolutions of 170 $\mu$m and 749 $\mu$m were obtained, respectively. Using the combined information from the TPC, ITC and VDET, a transverse momentum resolution of $\sigma(1/p_\mathrm{t}) = 0.6 \times 10^{-3}\,\mathrm{GeV}^{-1} \oplus 0.005/p_\mathrm{t}$ was achieved.

An electromagnetic calorimeter (ECAL) surrounded the TPC. This consisted of a lead-proportional wire chamber sampling device of thickness 22 radiation lengths which allowed the measurement of electromagnetic energy with a resolution for isolated leptons or photons of $\sigma(E)/E = 0.18/\sqrt{E} + 0.009$, where $E$ is the electromagnetic energy in GeV. The cylindrical superconducting coil which produced the axial magnetic field was situated outside the ECAL. The return yoke of the magnetic field, situated outside the coil, was fully instrumented to form a hadron calorimeter (HCAL) which was used to measure hadronic energy and also to serve as a muon filter. The energy resolution of this calorimeter was $\sigma(E)/E = 0.85/\sqrt{E}$ with $E$ the hadronic energy in GeV. Outside the iron structure, two double layers of streamer tubes, the muon chambers, provided two space coordinates for particles leaving the detector, thus improving the identification of muons. The luminosity was measured by downstream calorimeters covering small angles to the beam directions. The triggers for hadronic events were mainly based on the total ECAL energy deposited and muon track triggers. From comparison between independent triggers the trigger efficiency for Z decays was determined to be 99.7%, as described in [26].

Hadronic Z candidates were selected using the charged tracks. The events, taken when the apparatus was working well, were required to have at least five "good" tracks in the TPC with a summed energy greater than 10% of the summed energy of the electron and positron beams. A "good" track was defined as one with at least four reconstructed coordinates in the TPC and a polar angle $|\cos\theta| < 0.95$. In addition, it had to originate in a cylinder of radius 2 cm and length 20 cm centred at the known interaction point and parallel to the beam axis. This procedure was found to select a fraction of $97.48 \pm 0.02\%$ of the hadronic Z decays with negligible background [26].

## 2.2 Deuteron and Anti-deuteron Identification

Deuteron and anti-deuteron candidates were selected according to their momenta and specific ionization, $\mathrm{d}E/\mathrm{d}x$, estimated from the measured ionization samples in the TPC and normalised to the value expected from a minimum ionizing particle at the same polar angle. Figure 1 shows the $\mathrm{d}E/\mathrm{d}x$ values plotted against momenta for all tracks. Bands corresponding to electrons, pions, kaons, protons, deuterons and tritons and their anti-particles can be seen. The majority of the heavy ions in this sample are from spallation products due to secondary interactions.

Deuteron and anti-deuteron candidates with momenta measured in the range $0.55 < p < 1.0$ GeV/c were accepted if they had a value of $dE/dx$ more than 4.6 times minimum ionizing and a difference in the measured specific ionization from that expected for a deuteron or anti-deuteron, $E_s = dE/dx_M - dE/dx_E$, of less than 1.5. At smaller values of $dE/dx$, backgrounds from electrons and overlapping minimum ionizing tracks were encountered. The measured momentum range corresponds to a momentum range of $0.62 < p < 1.03$ GeV/c for candidates from the primary vertex, allowing for $dE/dx$ losses. A total of 72234 deuteron and 4994 anti-deuteron candidates remained after this selection. To suppress candidates from secondary interactions, the transverse distance of closest approach to the beam axis, $d_0$, of each candidate was required to be less than 0.4 cm, leaving 1788 deuteron and 51 anti-deuteron candidates. Further suppression of particles from secondary interactions was achieved by demanding that the tracks should not be reconstructed in a vertex remote from the primary interaction [27], leaving 1050 deuteron candidates and 50 anti-deuteron candidates. Figure 2 shows the distribution of $E_s$ for the anti-deuteron candidates, without the requirement that $E_s < 1.5$, in two ranges of the coordinate $z_b = z_{\text{track}} - z_v$. Here $z_{\text{track}}$ is the longitudinal position of the track at its closest approach to the beam axis and $z_v$ is the coordinate of the primary vertex in the event, reconstructed as described in reference [24]. It can be seen that $E_s$ is peaked strongly at zero, indicating that the majority of the selected tracks are compatible with anti-deuterons with few misidentified tracks from the very large number at lower values of $dE/dx$. Hence a pure sample had been obtained.

Deuterons are produced copiously as spallation products of the secondary interactions of primary particles in the material of the apparatus. Deuterons produced in the outer calorimeters, by such interactions, will be mislabelled as negatively charged anti-deuterons since they actually travel towards the interaction point while they are reconstructed assuming that they travel away from it. To determine the numbers of deuterons and anti-deuterons from primary interactions the measured values of $z_b$, defined above, were studied. The distribution of the measured values of $|z_b|$ for deuterons and anti-deuterons is shown in Fig. 3. Each distribution shows a peak at zero from tracks originating in the vicinity of the primary vertex with a background from tracks produced remotely from it. The background is flat for the anti-deuterons and decreases with $|z_b|$ for the deuterons. The decreasing non-primary background for deuterons can be understood as coming from spallation products produced relatively close to the interaction point while the approximately flat background for anti-deuterons originates from the distant production of spallation deuterons in the calorimeters, as explained above.

There were 11 anti-deuterons, from primary interactions, seen in the two bins with $|z_b| < 1.5$ cm with an estimated non-primary background of 0.2. This background was determined by extrapolating the flat background into the range $|z_b| < 1.5$ cm.

The number of deuterons from primary interactions is more difficult to determine because of the much larger non-primary background with an unknown shape (see Fig. 3). A possible component of this background could point towards the primary vertex due to spallation deuterons produced preferentially along the direction of the primary track. There were 196 deuterons seen in the range $|z_b| < 1.5$ cm. The non-primary background in the same range was estimated to be 142 by extrapolating a linear fit to the data in the range $1.5 < |z_b| < 22.5$ cm to lower $|z_b|$ i.e. a deuteron signal of $54 \pm 18$ events pointing towards the primary vertex, where the uncertainty is statistical. If the tighter selection $|d_0| < 0.1$ cm is made, the number of deuterons with $|z_b| < 1.5$ cm is reduced to 56 with a non-primary background of 40, i.e. $16 \pm 10$ primary events. The Monte Carlo simulation, described below, shows that this selection should reduce the deuteron and anti-deuteron reconstruction efficiency by a factor 0.66, which is compatible with the observed loss of two anti-deuterons (compared to $3.8 \pm 1.6$ expected). The difference between the numbers of deuterons obtained with the two different selections therefore shows that there is a large systematic uncertainty on the determination of this number. The number of deuterons could be compatible with the number of anti-deuterons within this large uncertainty. Due to this uncertainty only the anti-deuterons are used for quantitative measurements in the following.

## 2.3 Detection Efficiency for Anti-deuterons

The detection efficiency for anti-deuterons was calculated by Monte Carlo technique. Two difficulties needed to be overcome to accomplish this. Firstly, the Z Monte Carlo generators used to simulate the data did not include the production of anti-deuterons. Secondly, the simulation of the detector was based on GEANT [28] which allows tracking of deuterons and anti-deuterons, simulating ionization energy losses and multiple Coulomb scattering but not the losses of these ions due to nuclear interactions. In order to overcome these difficulties four sets of Monte Carlo simulated data were generated.

The first set was the standard Monte Carlo simulation of Z decays which uses the Lund Parton Shower Model (JETSET) [29] with the parameters set to those determined in [30]. The second set consisted of single pions in the same momentum and angular range as the anti-deuteron sample. The ratio of the fraction of the number of tracks reconstructed in the first set to that in the second set gives the probability of losing a charged particle within the jets of hadrons. The third set consisted of single anti-deuteron tracks generated within the standard Monte Carlo simulation. This was used to assess the efficiency for detection and reconstruction of anti-deuterons in the apparatus. The fourth set was generated by a simulation of anti-deuterons traversing the material of the apparatus in order to estimate the losses due to nuclear interactions.

The fraction of tracks lost within the jets of hadrons was measured to be $94.6 \pm 0.2\%$ using the first and second sets. In the third set the single anti-deuterons were generated with either a flat distribution in momentum or a distribution following the model of reference [22] and a cosine of the polar angle ($\cos\theta$) varying as $1 + \cos^2\theta$. The behaviour of these tracks in the apparatus was simulated (apart from nuclear interactions) and the same selection procedure applied to the simulated tracks as to the data. Figure 4 shows the reconstruction efficiency for the simulated anti-deuterons as a function of $\cos\theta$ (lower plot) and momentum (upper plot) for the sample generated as the momentum distribution of [22]. This shows that the reconstruction efficiency for anti-deuterons was $63\%$ on average. It was found to decrease to $59\%$ for the flat momentum distribution. Hence the reconstruction efficiency was taken to be the average of these i.e. $61 \pm 2\%$.

In the fourth Monte Carlo set, which was used to estimate the fraction of anti-deuterons lost by nuclear interactions, four different models were used to estimate the anti-deuteron nucleus total cross section which is not known from direct measurements.

- The first model employed the parameterisations of the inelastic $\bar{p}$ and $\bar{n}$ nuclear cross sections [31] and a geometric argument to combine these quadratically into the anti-deuteron cross section [32]. The cross section obtained was multiplied by two to allow for elastic scattering, the behaviour expected for a perfectly absorbing disc [33], as observed in deuteron-nucleus interactions [34]. This model gave the fraction of anti-deuterons lost in the material to be $18\%$.

- The second model is similar to the first but the $\bar{p}$ and $\bar{n}$ nucleus inelastic absorption cross sections computed from the parameterisations of [31] were added linearly to deduce an anti-deuteron nucleus inelastic cross section. Again the cross sections were doubled to allow for elastic scattering. This model gave $28\%$ for the fraction of anti-deuterons lost by nuclear interactions.

- The third model used the measured $\bar{p}$-nucleus total absorption cross sections of Ashford et al [37]. The assumption was made that the $\bar{n}$ and $\bar{p}$ total absorption cross section were the same and the two were summed to give an estimate of the anti-deuteron total absorption cross section. This method gave $17\%$ for the fraction of anti-deuterons lost by nuclear interactions.

- In the fourth model the measured $\bar{p}$-deuteron total absorption cross sections [38] were assumed to be the same as the total absorption anti-deuteron proton cross section at the same centre of mass energy. The anti-deuteron nucleus total absorption cross section was then obtained by multiplying this by the factor $A^{2/3}$ where $A$ is the atomic weight of the target nucleus. Such an $A$ dependence was roughly observed to describe the deuteron-nucleus measurements [34–36]. This method gave $24\%$ for the fraction of anti-deuterons lost by nuclear interactions.

Despite the crudity of these assumptions they give a roughly consistent picture of the total loss due to nuclear interactions. We take the mean of the four models as our best estimate of the fraction lost i.e. $22\%$. This value was for anti-deuterons with an angular distribution of $1 + \cos^2\theta$ and a momentum distribution following the model of reference [22]. Using a flat angular distribution or a flat momentum dependence led to values of 23 and $20\%$ for the fraction lost. From this variation together with that from the different computed cross sections we take the uncertainty on the fraction lost to be $\pm 6\%$ i.e. the fraction lost is $22 \pm 6\%$.

It can be seen from Fig. 4 that the reconstruction efficiency for anti-deuteron detection decreases for $|\cos\theta| > 0.8$. In addition, the efficiency goes to zero at momenta smaller than 0.6 GeV/c. These inefficient regions can be understood from a combination of two effects. Firstly, the number of TPC hits becomes too small to reconstruct the $dE/dx$ at low values of transverse momenta and, secondly, low momentum anti-deuterons are lost at oblique angles, coming to the end of their range in the material of the apparatus. The efficiency also decreases to zero for momenta above $1.03\,\text{GeV/c}$ since $dE/dx$ becomes less than the minimum selected in this analysis (4.6 times minimum ionizing). The overall efficiency, made up of the probability of the track to be reconstructed in a hadronic Z decay ($94.6 \pm 0.2\%$), the probability of the track to be reconstructed as an anti-deuteron ($61 \pm 2\%$) and the probability to escape nuclear absorption ($78 \pm 6\%$), is $\epsilon = 45 \pm 4\%$.

# 3 Results

The number of anti-deuterons per hadronic Z decay, $R_{\bar{D}}$, from the observed anti-deuteron signal is

$$R_{\bar{D}} = \frac{10.8 \pm 3.3}{\epsilon \, N_Z} = (5.9 \pm 1.8 \pm 0.5) \times 10^{-6}, \quad (2)$$

in the momentum range $0.62 < p < 1.03$ GeV/c ($\Delta p = 0.41$ GeV/c) and $|\cos\theta| < 0.95$, where the first uncertainty is the statistical and the second the systematic error. Here $N_Z = 4.07 \times 10^6$ is the observed number of hadronic Z decays and $\epsilon = 45 \pm 4\%$ is the anti-deuteron detection efficiency, (see Section 2.3). The systematic error represents the uncertainty on this efficiency. The measured value of $R_{\bar{D}}$ is compatible with the upper limit at 95% confidence level of $8 \times 10^{-6}$ set on the rate of anti-deuteron production by the OPAL Collaboration [21] in a slightly wider momentum range ($0.35 < p < 1.1$ GeV/c). It follows from equation 2 that the value of the inclusive cross section for anti-deuteron production is

$$\frac{1}{\sigma}\frac{d\sigma}{dp} = \frac{R_{\bar{D}}}{\Delta p} = (1.4 \pm 0.4 \pm 0.1) \times 10^{-5} \, (\text{GeV/c})^{-1} \quad (3)$$

and the invariant cross section averaged over the whole angular range

$$\frac{1}{\sigma} E \frac{d^3\sigma}{d^3 p} = C \frac{E}{p^2} \frac{1}{\sigma} \frac{d\sigma}{dp} = (3.7 \pm 1.1 \pm 0.3) \times 10^{-6} \, \text{GeV}^{-2} \text{c}^{-3} \quad (4)$$

where $E$ and $p$ are the energy and momentum of the anti-deuterons at the centre of the momentum range and $C = 0.0859$ is the reciprocal of the total sensitive solid angle allowing for the fact that the angular distribution is approximately $1 + \cos^2\theta$.

# 4 Discussion of the Results

Figure 5 shows the measured value of the inclusive anti-deuteron cross section compared to the model of reference [22] which predicts suppression of anti-deuteron production in $e^+e^-$ collisions. The measurement presented here indicates that such suppression is overestimated in the model.

From these cross sections the coalescence parameter, $B_2$, defined according to equation 1 with $A = 2$, for anti-deuteron production in Z decays is

$$B_2 = \frac{\frac{1}{\sigma}\frac{E_{\bar{d}} d^3\sigma_{\bar{d}}}{d^3 P}}{\left(\frac{1}{\sigma}\frac{E_{\bar{p}} d^3\sigma_{\bar{p}}}{d^3 p} F\right)^2} = (3.3 \pm 1.0 \, (\text{stat.}) \pm 0.8 \, (\text{sys.})) \times 10^{-3} \, \text{GeV}^2. \quad (5)$$

where the invariant cross section for anti-proton production is taken to be $0.044 \pm 0.004$ GeV$^{-2}$ in the range $0.31 < p < 0.52$ GeV [39] and $F$ is the fraction of anti-protons from direct production, excluding those from weak decays. In the coalescence model, only anti-nucleons produced directly from the source can form anti-deuterons. The number of $\bar{p}$ from direct production is taken to be a fraction of 0.76 of the number observed. This is estimated from the PYTHIA Monte Carlo [40] and is thought to be accurate to about $10\%$ due to the uncertainty in the strangeness suppression parameter in this model. Hence there is an overall theoretical uncertainty of about $20\%$ in the determination of $B_2$. The systematic error (second error) on $B_2$ includes this theoretical error as well as the uncertainty on $\epsilon$ and the uncertainty in the invariant anti-proton proton cross section.

Figure 6 shows this measured value of $B_2$ compared with the measurements of $B_2$ obtained from other experiments using "elementary" projectiles i.e. pA [5,17,18], pp [13,14] and $\gamma$p [19] collisions and those in very heavy ion collisions. Here, $B_2$ has been calculated for the ISR pp data from the cross sections given in [13–15] taking the direct fraction to be 0.75 which was estimated from the PYTHIA program [40]. The limit on $B_2$ from the OPAL data in $\mathrm{e^+e^-}$ collisions is also shown. This limit was calculated from their anti-deuteron rate limit and anti-proton cross sections in the continuum [39] with a direct fraction calculated to be 0.76. The data are restricted to inclusive deuteron and proton production for the Bevelac and AGS data to avoid threshold effects in anti-deuteron production. The ion-ion data are restricted to very heavy ions to reduce the sensitivity to A dependent effects, and to the most central data, since the measurements show dependence on centrality [4,10]. The data from the Bevelac, at which energy the A dependence is weak, are the Ne-Au measurements of [17]. The data from the AGS are the Au-Pt measurements of the E886 experiment [5] and the Au-Au data of experiments E864 [6] and E896 [7], the data from the SPS are the Pb-Pb measurements of NA44 [9], NA49 [10] and NA52 [11] and the data from RHIC are from the Au-Au measurements of the STAR and PHENIX collaborations [3,4]. It can be seen that the values of $B_2$ are smaller in both heavy ion and $\mathrm{e^+e^-}$ collisions than those measured in collisions involving "elementary" projectiles i.e. protons or photons. However, the suppression in heavy ion collisions is more marked than in $\mathrm{e^+e^-}$ collisions.

## 5  Conclusions

Deuteron and anti-deuteron production has been observed in $\mathrm{e^+e^-}$ collisions at the Z resonance energy. The number of anti-deuterons per Z decay was measured to be $(5.9\pm1.8\pm0.5)\times10^{-6}$ in the momentum range 0.62 to 1.03 GeV/c. The data were used to determine that the coalescence model parameter, $B_2 = (3.3\pm1.0\pm0.8)\times10^{-3}$ GeV$^2$ in $\mathrm{e^+e^-}$ annihilations at the Z resonance. This is smaller than that measured in hadronic and photonic collisions with protons, indicating the suppression of the coalescence process in $\mathrm{e^+e^-}$ collisions. However, the measured value of the inclusive cross section $\frac{1}{\sigma}\frac{d\sigma}{dp} = (1.4\pm0.4\pm0.1)\times10^{-5}\,\mathrm{GeV}^{-1}$ is higher than the prediction of the model of reference [22].

## 6  Acknowledgements

It is a pleasure to congratulate and thank our colleagues from the accelerator divisions at CERN for the successful operation of LEP. We wish to thank also G. Gustafson and J. Nystrand for useful discussions. We are indebted to the engineers and technicians at our home institutes without whose dedicated help this work would not have been possible. Those of us from non-member states wish to thank CERN for its hospitality and support.

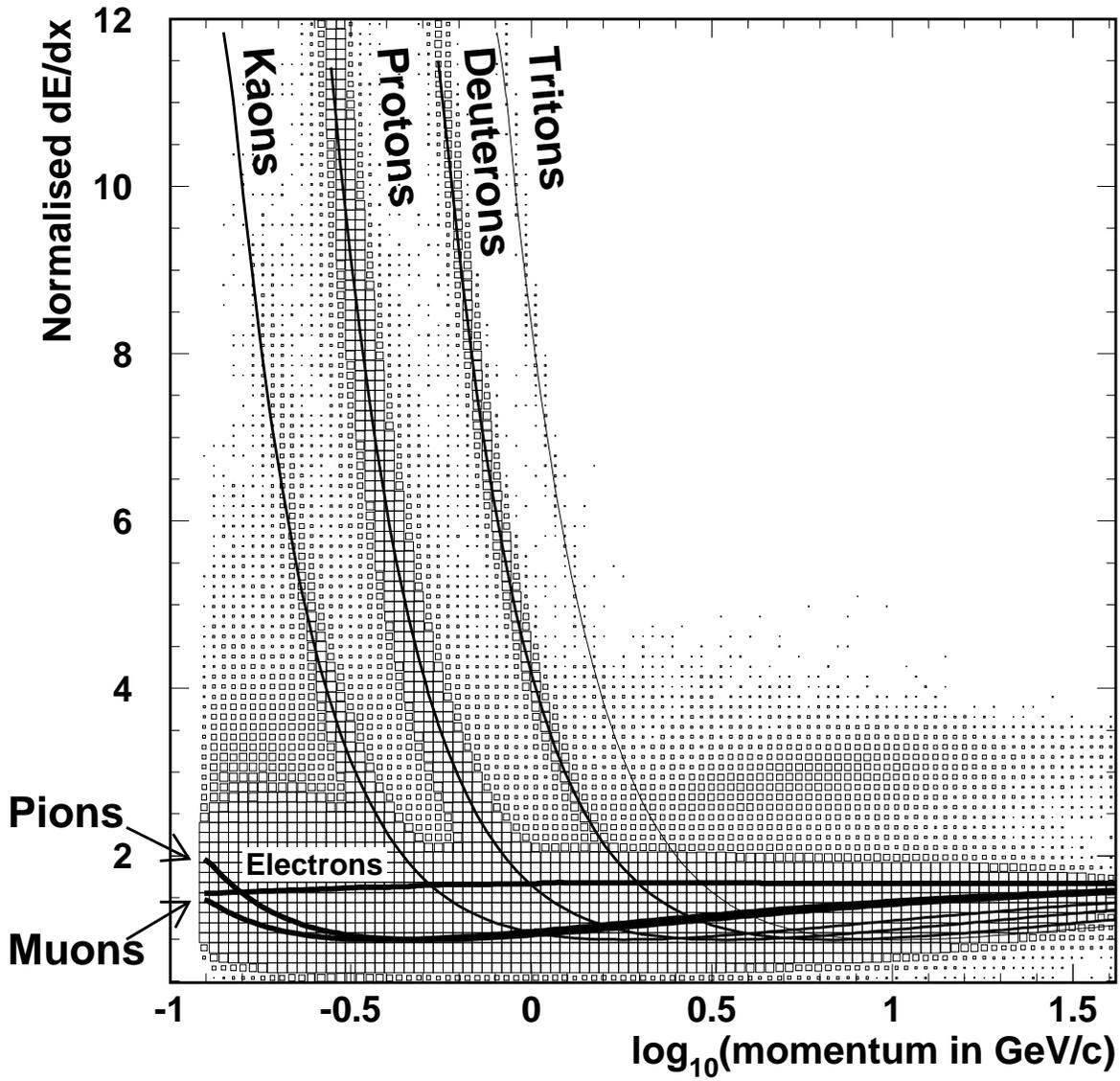

Figure 1: The value of $dE/dx$, normalised to that of a minimum ionizing particle, as a function of the logarithm of the momentum for a sample of all tracks. Bands corresponding to electrons, pions, kaons, protons, deuterons and tritons can be seen.

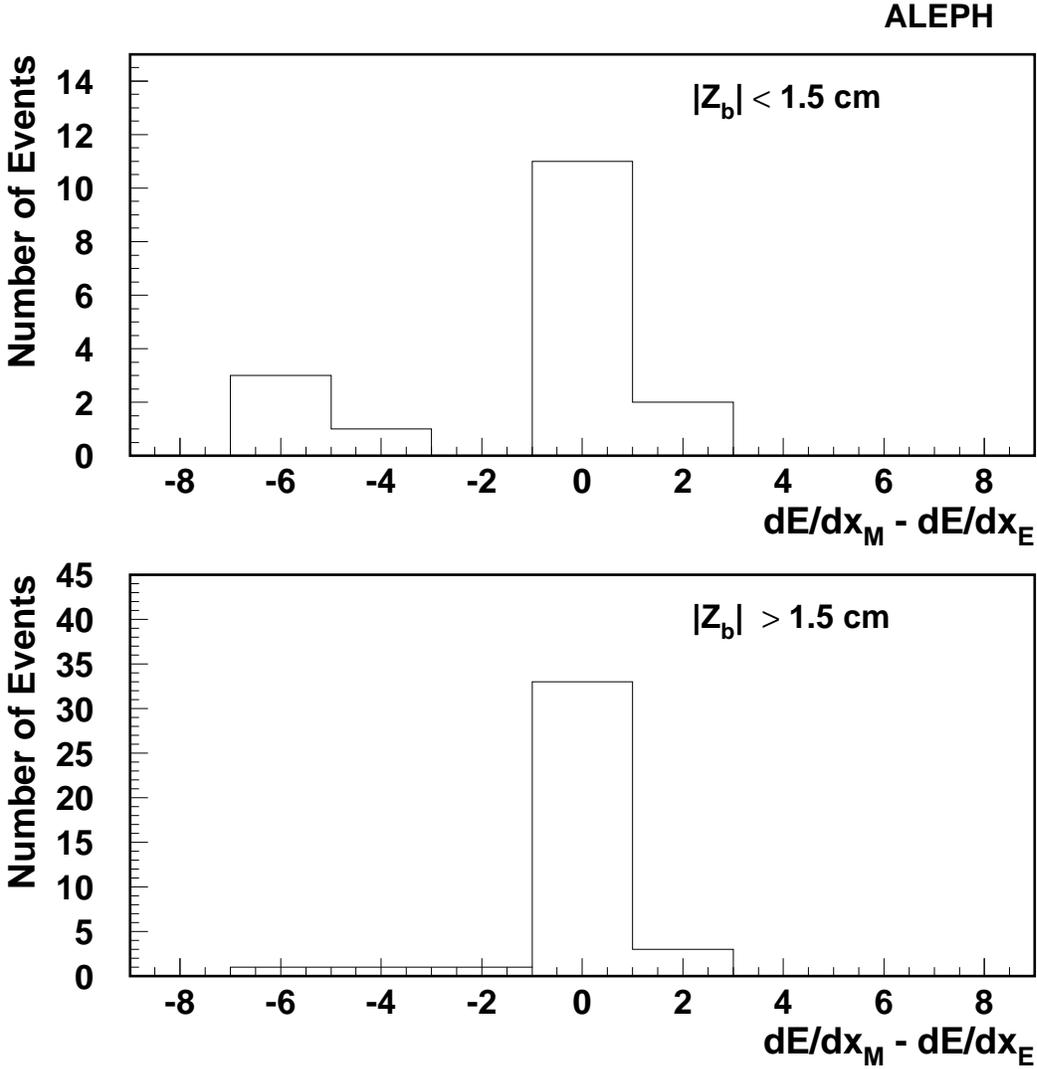

Figure 2: The distribution of the values of $E_{\mathrm{s}} = \mathrm{d}E/\mathrm{d}x_{\mathrm{M}} - \mathrm{d}E/\mathrm{d}x_{\mathrm{E}}$, in two ranges of $z_{\mathrm{b}}$, the longitudinal position of the track relative to the vertex at the closest approach to the beams, for all anti-deuteron candidates without the requirement that $E_{\mathrm{s}} < 1.5$. Here $\mathrm{d}E/\mathrm{d}x_{\mathrm{M}}$ and $\mathrm{d}E/\mathrm{d}x_{\mathrm{E}}$ are the measured and expected values of $\mathrm{d}E/\mathrm{d}x$ for anti-deuterons, each normalised to the value for a minimum ionizing particle. The tracks in the upper plot are selected to be in the signal region, as explained in the text, while those in the lower plot are in the nuclear spallation region.

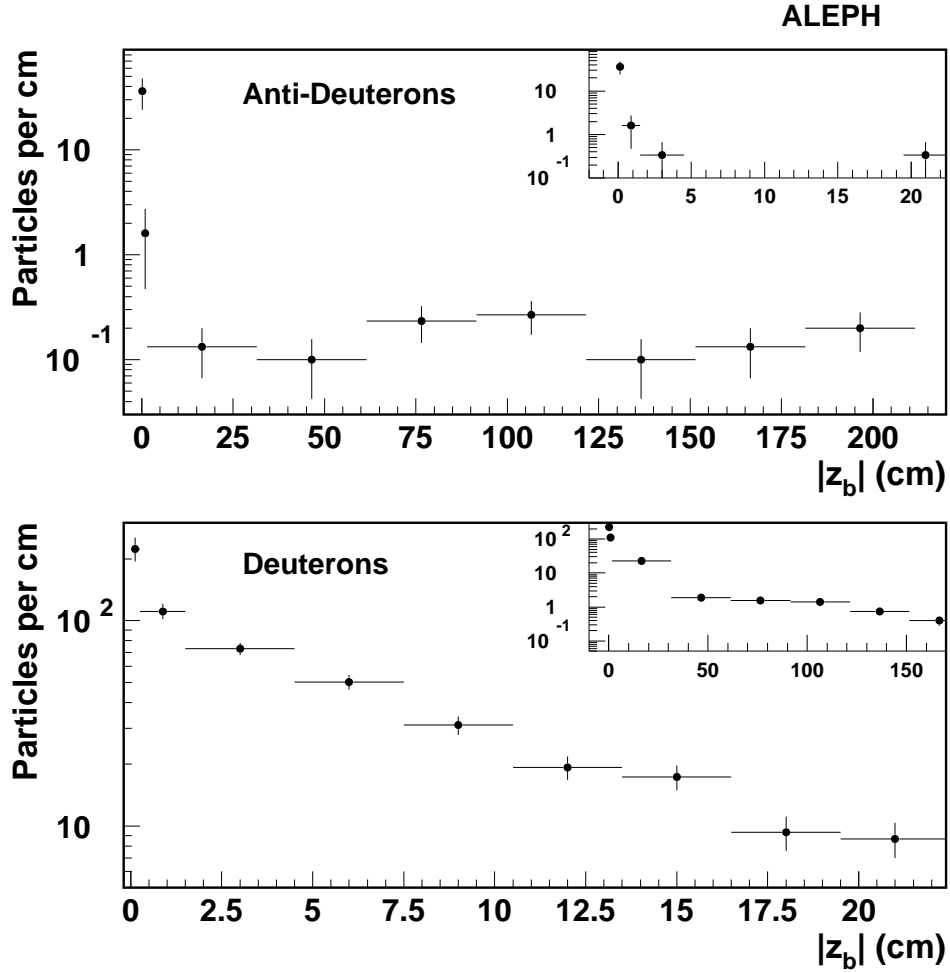

Figure 3: The distribution of the number of particles per cm of measurement interval against the longitudinal coordinate at the closest approach of the track to the beam line, $z_b$, for the final samples. The peak near zero includes primary events from $e^+e^-$ annihilation collisions while the background away from zero is due to spallation products. The vertical bars show the statistical errors and the horizontal error bars represent the measurement intervals. The first two intervals cover the ranges $0 < |z_b| < 0.25$ cm and $0.25 < |z_b| < 1.5$ cm and the remainder are each 30 cm wide or 3 cm wide on the plots covering the range $0 < z_b < 22.5$ cm.

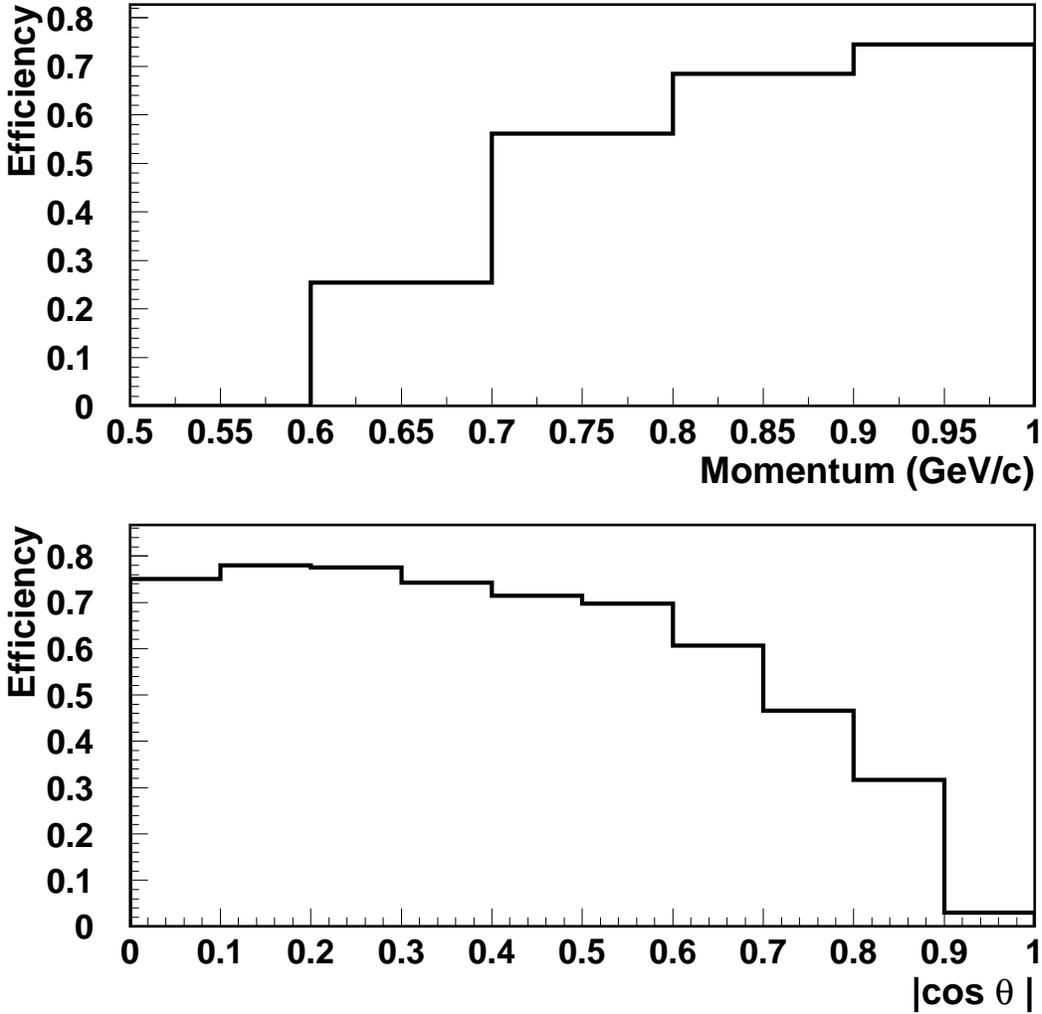

Figure 4: The anti-deuteron reconstruction efficiency as a function of the generated momentum (upper plot) and cosine of the polar angle (lower plot), as determined from the single particle Monte Carlo simulation with the momentum distribution of [22] (see text). The reconstruction efficiency includes that of a particle in a Z decay event (94.6%).

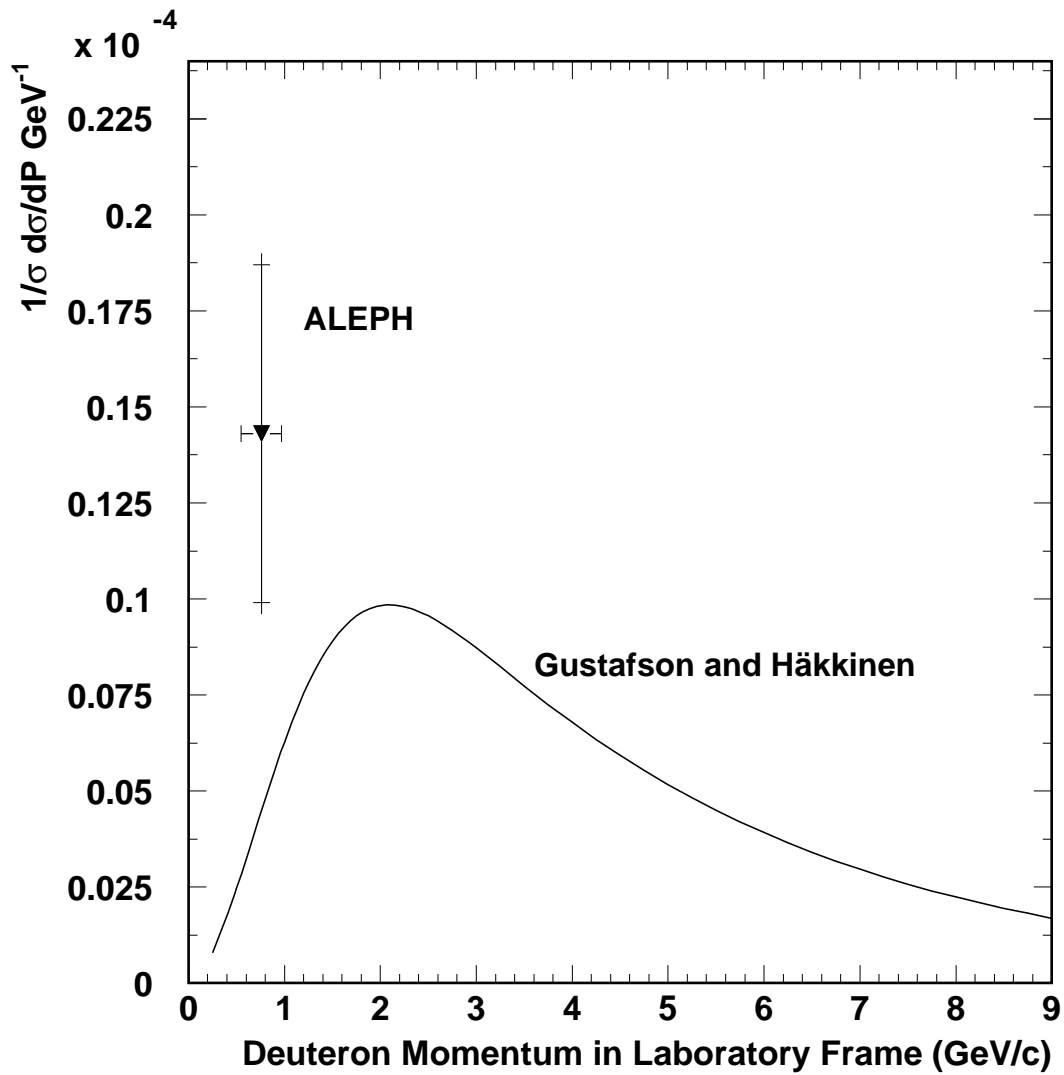

Figure 5: The predicted inclusive anti-deuteron cross section in $e^+e^-$ collisions at the Z resonance from the model of [22] (solid curve) compared to the value measured here shown by the point labelled ALEPH. The inner error bar shows the statistical uncertainty and the outer error the total uncertainty given by the sum in quadrature of the statistical and systematic errors. The horizontal bar shows the measurement interval.

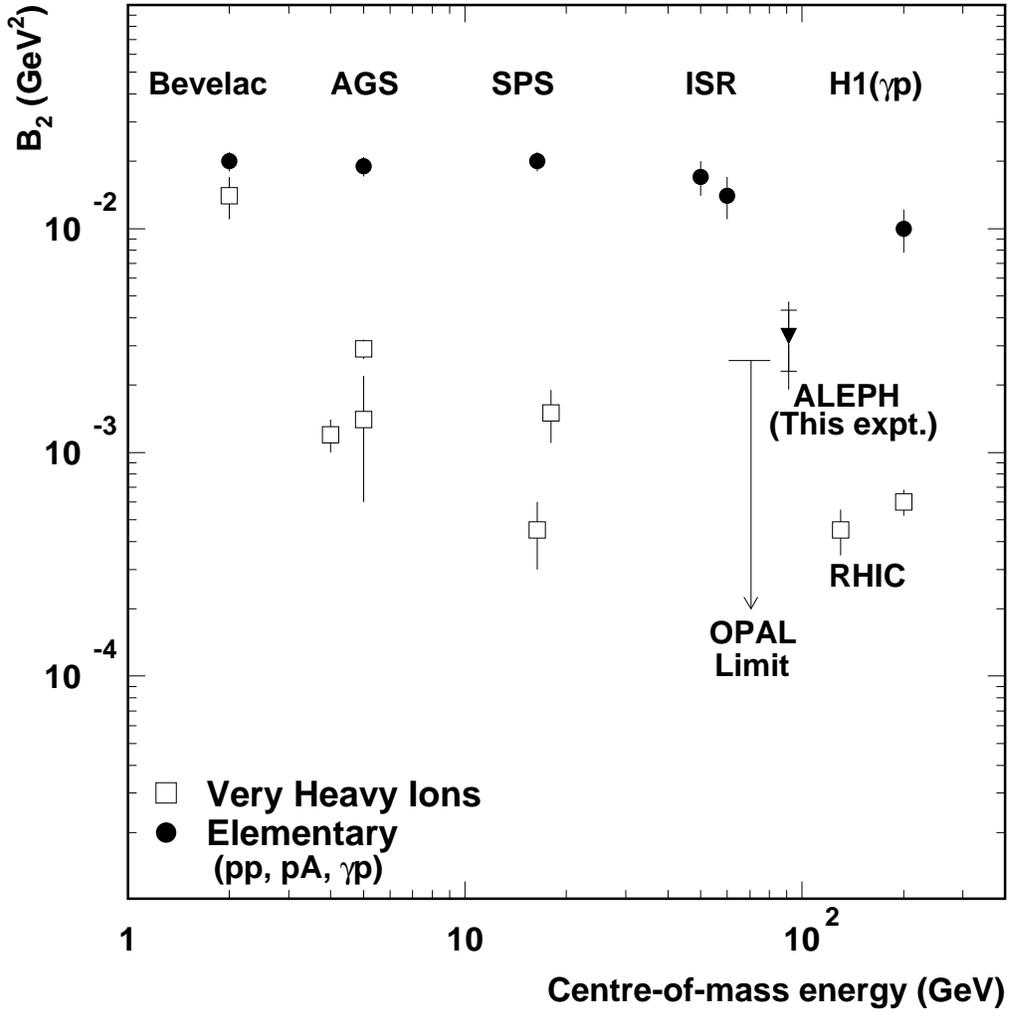

Figure 6: The values of the coalescence parameter, $B_2$, measured in very heavy ion collisions and with collisions between more elementary hadrons compared to the value in $e^+e^-$ collisions from this experiment (labelled ALEPH) and the limit from OPAL [21]. The sources of the "elementary" and heavy ion data are given in the text.

16